\documentstyle[prl,aps,preprint]{revtex}
\begin{document}
\bibliographystyle {prsty}
\preprint{WIS--9.--PH}
\draft
\title{Inter-Edge Interaction in the Quantum Hall Effect}

\author{Yuval Oreg and Alexander M. Finkel'stein \cite{Lan}}

\address{ Department of Condensed Matter, The Weizmann Institute of Science,
Rehovot 76100, Israel}
\date{\today}
\maketitle

\begin{abstract}

We consider effects of the interaction between electrons drifting along
the opposite sides of a narrow sample under the conditions of the quantum
Hall effect. A spatial variation of this interaction leads to backward
scattering of collective excitations propagating along the edges.
Experiments on propagation of the edge modes in samples with constrictions
may give information about the strength of the inter-edge electron
interaction in the quantum Hall regime.

\end {abstract}


\def\l{\left(}
\def\r{\right)}
\def\beqa{\begin{eqnarray}}
\def\eeqa{\end{eqnarray}}
\def\beq{\begin{equation}}
\def\eeq{\end{equation}}
\def\barr{\begin{array}}
\def\ear{\end{array}}

A potential, which confines the two-dimensional electron gas inside a
sample, leads to the formation of the so-called edge states near the
boundaries of the sample in the presence of a magnetic field
 \cite{QHE:Halperin82}.  When the chemical potential lies in a gap of the
bulk levels the role of the edge states becomes dominant and many
experiments in the quantum Hall regime may be interpreted by transport of
the edge electrons \cite{QHE:Beenakker91,QHE:Haug93} or by propagation of
collective edge modes
\cite{FQHE:Wassermeier90,QHE:Ashoori92,QHE:Talyanskii92}.  The collective
edge excitations are analogous to the edge magnetoplasmon modes
 \cite{QHE:Volkov88}, but under the conditions of the quantum Hall effect
the specifics of the energy spectrum of the electrons near the edges makes
these modes rather peculiar \cite{FQHE:Wen92}.

In this work we discuss effects of the interaction between electrons
drifting along opposite sides of the sample. These effects may be observed
at time-resolved transport phenomena in samples with non-constant
effective width, e.g., in a strip with constrictions. We show that the
variation of the inter-edge interaction due to the constrictions leads to
backward scattering of the collective excitations propagating along the
opposite edges. The scattering of waves from one boundary of the sample to
the other one is not related with the direct hopping of electrons between
the edges.  Due to a long-distance electron--electron interaction this
scattering may happen when the electron hopping from one side of the
sample to the opposite one is completely forbidden. For a particular
sample with an extended constriction, which acts as a semi-transparent
cork, one may observe oscillations in the transparency of the sample as a
function of the frequency of the edge wave. The magnitude of the effect
gives information about the strength of the inter-edge interaction, while
the period of the transparency oscillations provides one, for a given geometry
of the sample, with the value of the velocity of the collective edge
modes.  In samples with rough boundaries the backward scattering of waves
on random inhomogeneities of the boundaries opens a channel for relaxation
of the electrons at the edges.  This mechanism will be discussed in the
final part of the paper.

The strength of the inter-edge interaction depends on the particular
electrostatics of the sample. We intend to present here only a general
idea of studying of the inter-edge electron-electron interaction. For this
purpose we shall take as a base a simple picture of an abrupt potential
near the edges when the transition between the filled and empty states is
sharp.  We believe, that in such systems the phenomena related with the
inter-edge electron coupling will be more pronounced compared with the
systems where gate-confined electron density slowly varies in the lateral
direction.  For simplicity we restrict the present discussion to
situations when there is only one pair of edge states, i.e., when the
filling factor of the lowest Landau level $\nu=1/n$ where $n$ is odd
\cite{QHE:Mac90}. Bellow we concentrate on the case $\nu=1$ and we will
comment upon the fractional filling in the concluding part.

To begin with, we consider the effect of the electron-electron interaction
on the Hall conductance. It will be shown that despite the fact that the
diamagnetic current along the boundaries is affected by the interaction,
the static limit of the Hall conductance is still quantized when backward
hopping of electrons from one edge to the opposite one is absent. This
point was already discussed in Ref. \cite{EDGE:Wen91}, but now it will be
proven in a rather general way.

We concentrate bellow on excitations with energy much smaller than
the cyclotron frequency. Therefore, the transversal motion of electrons can be
excluded by means of the adiabatic approximation. Then, only the
longitudinal motion along the strip remains essential, and finally
one comes to the picture of effectively one--dimensional fermions.
The momentum quantization along a conducting strip of a quantum Hall
device $k_n=2\pi n /L_x \; (n=0,\pm1,\pm2,...;\;  L_x$ is the length of
the strip along the drift direction) leads to quantization of the center
of the orbit of magnetized electrons according to $y_n=l_c^2 k_n$, where
$l_c=\sqrt {\hbar c /eB}$ is the magnetic length. The momenta $k_{u}$ and
$k_{l}$ of the last occupied states at the ``upper'' and ``lower'' edges
correspond to the $\pm k_F$ of the 1-d electron gas, while the drift
velocity of the edge electrons is the analog of the bare Fermi velocity.
The main difference of the effective one-dimensional theory at hand from
the conventional 1-d electron gas is the absence of time inversion
symmetry: particles which are moving in opposite directions are spatially
separated and may have different velocities. For free electrons the
quantization of the conductance can be easily obtained in this picture
\cite{QHE:Mac84}.

Let us now discuss the quantization of $\sigma_{xy}$ in the presence
of electron--electron interactions. Due to the mutual coupling of
electrons drifting along the edges their velocity is renormalized.  A
convenient and economical way to describe the velocity renormalization is
to derive the current operators using the continuity equation.  When
hopping of electrons from one edge to the other is absent the species of
electrons on each boundary are well defined, and we can apply the
continuity equation for each edge current separately:

\beq
\label {Eq:Current-operator}
J_{u,l}(p)=\frac{i}{p} \frac{d}{dt} e \rho_{u,l} (p) =
\frac{e}{ \hbar p} \left[ H,\rho_{u,l}(p) \right],
\eeq

where $H$ is the Hamiltonian of the system, and
 the operators $\rho_{u,l}(p)=\frac{1}{\sqrt{L_x}}\sum_{k
\approx k_{u,l}} a^{\dagger}_{k+p} a_k$ describe the creation of
 charge density excitations  on the edges.
For the part of the Hamiltonian that describes
 excitations with energy less than the cyclotron frequency the
 bosonized representation \cite{FQHE:Wen92,EE1D:Solyom79} will be exploited:
\beqa
\label{Eq:H}
H= \pi\sum_{p} v_u  \rho_u( p) \rho_u(-p)
   + \pi\sum_{p} v_l  \rho_l(-p) \rho_l( p) + \nonumber \\
     \frac{1}{2} \sum_{p} W_{u}(p) \rho_u( p) \rho_u(-p)
   +    \frac{1}{2} \sum_{p} W_{l}(p) \rho_l(-p) \rho_l( p) +           \\
 \sum_{p} U(p) \rho_l(p) \rho_u(-p)+ \mbox{Anharmonic terms}. \nonumber
 \eeqa
 Here the terms with $v_u$ and $v_l$ represent the energy spectrum of
magnetized electrons linearized near the edges, while the nonlinear part
of this spectrum can be written in the bosonization technique in the form
of an anharmonic interaction \cite{EE1D:Schick68}.  The amplitudes
$W_{u,l}$ describe the intra-edge Coulomb interaction which is responsible
for the enhancement \cite{QHE:Volkov88,FQHE:Wassermeier90} of the velocity
of the edge modes. The $U$-term describes the inter-edge electron-electron
interactions. In fact our consideration holds for any Hamiltonian
$H\left\{ \rho \right\}$ describing the edge-states physics by means of
a fancional of the $\rho$- operators.
 The $\rho_{u,l}$-operators in Eq. (\ref{Eq:H}) have
the standard 1-d commutation relations \cite{FQHE:Wen92,EE1D:Mattis65}:
\beq
 \label{Eq:cr}
 \renewcommand{\arraystretch}{1.5}
\barr{c}
 \left[\rho_u(-p),\rho_u(p^\prime)\right]=
 \left[\rho_l(p),\rho_l(-p^\prime)\right]=
\frac{p}{2 \pi}\delta_{p,p^\prime};\\
\left[\rho_l(p),\rho_u(-p^\prime)\right]=0.
\ear
\eeq
For operators commuting like that,  performing commutation
is equivalent  to differentiation, i.e.,
$
\left[ F \left\{ \rho_{u,l} \right\},  \rho_{u,l}(q) \right]= \pm \frac
{q}{2\pi} {\partial F\left\{\rho_{u,l}\right\}} /
{\partial \rho_{u,l}(-q)}$.
Therefore we can rewrite $J_{u,l}$ in Eq. (\ref{Eq:Current-operator})
as
\beq
\label{Eq:current=diffrential-energy}
J_{u,l}(p)=\pm \frac{e}{2\pi \hbar} \frac{\partial H}
{\partial \rho_{u,l}(-p)}.
\eeq
On the other hand, by definition, the chemical potential of the edges is
\beq
\label{Eq:Chemical-potential}
\mu_{u,l}(p)=\frac{\partial H} {\partial \rho_{u,l}(-p)}.
\eeq
Thus, for the total current $I=J_u+J_l$ we obtain
\beq
\label{Eq:I(mu)}
I(p)= \frac{e}{2\pi \hbar}\left( \mu_u(p)-\mu_l(p) \right).
\eeq
This gives the quantization of the static Hall conductance in units
 $\frac{e^2}{2 \pi \hbar}$ (in fact the quantization holds for any
 $p$, i.e., locally).  It should be emphasized that the electron
interactions affect both the currents $J_{u,l}$ and the chemical
potentials $\mu_{u,l}$. However the structure of these corrections is such
that the total current is changed in the same way as the difference of the
potentials. As a result, these corrections proved to be canceled in the
ratio that determines the conductance. Provided that there is no
electron hopping between the edges, this fact is obtained here relying only on
the
representation of the Hamiltonian as a functional of the $\rho$-operators.
Such representation is not well defined, however, when the density of
states at the Fermi energy is singular.  For that reason direct
application of the above consideration to a system with alternating strips
of compressible and incompressible liquids \cite {QHE:Chklovskii92} is not
possible.

Thus, the Hall transport in the static limit does not provide us with any
information on the electron interactions. With a purpose to reveal an
effect of the interaction between electrons on different edges let us
consider the propagation of the edge modes in a inhomogeneous system. The
general form of the bilinear part of the Hamiltonian (\ref{Eq:H}) is
\beq
\label{Eq:bilinear}
H= \frac{1}{2} \int V_u(x,y) \rho_u(x) \rho_u(y) dx dy +
\eeq
$$
     \frac{1}{2} \int  V_l(x,y) \rho_l(x) \rho_l(y) dxdy +
       \int       U(x,y) \rho_u(x) \rho_l(y)  dx dy
$$
In order to diagonalize the Hamiltonian (\ref{Eq:bilinear}) we will write
$\rho_{u,l}$-operators as
\beq
\label{Eq:BT}
 \renewcommand{\arraystretch}{1.5}
\barr{ccc}
\rho_u(x)&=&\sum_n \rho_{I }(-n) \eta_{-n}(x)- \rho_{II}(-n) \chi_{n}(x)\\
\rho_l(x)&=&\sum_n \rho_{II}( n) \eta_{-n}(x)- \rho_{ I}( n) \chi_{n}(x),
\ear
\eeq
where $\rho_{I}$ and $\rho_{II}$ are new operators still satisfying the
commutation relations of Eq. (\ref{Eq:cr}) in which $\rho_u \rightarrow \rho_I$
and
$\rho_l \rightarrow \rho_{II}$.
The transformation (\ref{Eq:BT}) is an inhomogeneous  variant
 of the Bogoliubov transformation
similar to the one used  in the theory of superconductivity
\cite{SC:De-Gennes64}.
This transformation diagonalizes the Hamiltonian (\ref{Eq:bilinear})
if the conditions \mbox{
$ \left[H,\rho_{I }( n) \right]= v_{I }( n) n \rho_{I }( n)$} and \mbox{
$\left[H,\rho_{II}(-n) \right]= v_{II}( n) n \rho_{II}(-n)$}
are fulfilled.
To derive  the equations for the eigenfunctions of the
wave modes $\eta$ and $\chi$, we rewrite
commutation relations (\ref{Eq:cr}) in the coordinate form
and  calculate
the commutators $\left[H,\rho_{u,l}(x) \right]$. Then,
replacing $\rho_{u,l}$ by means of $\rho_{I,II}$  we obtain
\beq
 \label{eq:Bog-eq}
 \renewcommand{\arraystretch}{1.5}
\begin{array}{ccc}
 \omega_n\eta_n(x)&=&i\int dy \left[\partial_x V_u(x,y)\eta_n(y)-\partial_x
U(x,y)  \chi_n(y) \right]\\
 \omega_n\chi_n(x)&=&i\int dy \left[\partial_x U(y,x)  \eta_n(y)-\partial_x
V_l(x,y)\chi_n(y) \right],
\end{array}
\eeq
where $\omega_n=v_I(n)n$, and  a pair of equivalent equations  with
 $\omega_{-n}=v_{II}(n)n$.
In the presence of the inter-edge interaction the eigenmodes are not
localized anymore near one of the edges, but are combined from
excitations which are located on both sides. These modes can still
be classified as left and right movers. In the homogeneous case
when the potentials $V_{u,l}(x,y)$ and $U(x,y)$ depend only on
$(x-y)$ Eqs. (\ref{eq:Bog-eq}) reproduce correctly the well known
solution of the Tomonaga-Luttinger model \cite{EE1D:Mattis65}, which
gives two modes propagating with the velocities $ v_{I,II}=\pm
\frac{1}{4\pi} \left\{ V_u(k)-V_l(k) \pm \sqrt{ (V_u(k)+V_l(k))^2-
4U(k)^2} \right\}$, where $V_{u,l}(k)$ and $U(k)$ are the Fourier
transforms of the potentials. The amplitudes $V_{u,l}(k)$ may have  a
logarithmic dependence on $k$, if the  Coulomb interaction is not
efficiently screened.

  Without loosing generality it will be assumed bellow that all effects of
inhomogeneity are only due to $U(x,y)$.  When the wavelength of an
eigenmode is much shorter than the characteristic length on which the
potential $U$ changes, the adiabatic approximation can be applied. In that
case the propagating modes adjust themselves to the local value of the
interaction $U$ and no reflection occurs.  The opposite situation, for
which the sudden approximation is valid, occurs when the wavelength of the
eigenmode is larger than the region where the potential $U$ alters. It can
be realized either in a sample with a sudden narrowing of the conducting
strip, or in a sample, which is partially covered by a metallic gate or by
 a material with a different dielectric constant. We will model this
 situation by a potential $U(x,y)$ that vanishes at $ x,y<0 $, while for
$x,y >0$ we take $U(x,y)=U \delta(x-y)$ assuming that the
 characteristic length of the action of the potential $U(x,y)$ is
  shorter than the wavelengths of the eigenmodes.  Consider now a mode
propagating along the upper edge to the right.
  When the incident wave reaches the region of the inter-edge interaction
a backward wave is excited on the lower edge, since in the presence of
inter--edge coupling the eigenmodes are built from waves which are located
on both sides.  By matching the wave solutions for the
semi-infinite strips, we find the transmission coefficient of the incident
wave $ T= \left\{ 1+r(1+r) \right\}^{-1}, $ where $r=\left[(V_u-2 \pi
v_I)/U\right]^2$.  For the case of symmetric boundaries, when $V_u=V_l=V$,
the velocity $v_I= \frac{1}{2 \pi}\sqrt{V^2-U^2}$, and  for $U \ll  V$ we
 obtain
 \beq T= 1-\frac{1}{2} \left( \frac{U}{ V} \right)^2 \eeq

Another geometry that we consider  is a sample in which inter-edge
interaction acts inside a constriction of length $L_{int}$:
 $U(x,y)=U \delta (x-y) \mbox{ for } 0<x,y<L_{int}$ and is equal to
  zero otherwise.  In such a free-interacting-free (FIF) junction we
  find that due to the multiple  back and forth reflections the
transmission coefficient oscillates according to:
  \beq
\label{Eq:oscillations} T(\omega)=\frac{1} {1+|\frac{U}{
V}|^2\sin^2(2 \pi \frac{\omega}{V}L_{int})}.  \eeq
 Here it was
assumed again that $V_u=V_l=V$ and $U \ll  V$.  These oscillations
resemble the oscillations of the differential resistance of a
superconducting junction, known as the Tomasch oscillations \cite
{SC:Tomasch65}.  (We may add in that respect, that the reflection of
waves studied here by means of the Bogoliubov transformation has some
similarity with the Andreev reflection \cite {SC:Andreev64}.) An
experiment on a  FIF junction may provide us with information about
the magnitude of the inter-edge interaction.

Now consider the reflection of edge waves due to inhomogeneities of
the boundaries.  In  resonance experiments on quantum wires or
annulus samples this
reflection mechanism can determine the width of the resonance.  A
random variation of the shape of the boundaries creates a random
 sequence of  potentials, which are similar to the potential that was
studied at the derivation of Eq. \ref{Eq:oscillations}. Assuming that
the typical length of the inhomogeneities, $L_{int}$,  is smaller
than the wavelength, we model this situation by the
inter-edge potential \beq \label{Eq:random-potential} U(x,y)=\sum_i
U_i(x-a_i)\delta(\frac{x-y}{L_{int}}), \eeq where $a_i$ are the
locations of the inhomogeneities and $U_i$ are some short range
potentials.

The propagators of the edge waves will be  defined by ${\cal
D}_{u(l)}(x-y,t-t^\prime) = -i \left< T (\rho_{u(l)}(x,t)
\rho_{u(l)}(y,t')) \right>$. In the $q,\omega $ representation the
free propagators are given by:  \beq {\cal  D}^0_u=\frac{
 q}{\omega-v_Iq+i\delta  sign(q)}; \;\; {\cal
D}^0_l=\frac{-q}{\omega+v_{II}q-i\delta sign(q)}. \eeq In the
 presence of the potential (\ref {Eq:random-potential}) the averaged
propagators ${\cal D}_{u,l}$ can be found by averaging over $a_i$
like in the case of electrons scattered by random impurities
\cite{RFS:AGD63}.  For the self energy $\Sigma$ of the Dyson equation
we obtain in this way
 \beq \label{Eq:sigma} \Sigma_{u(l)}= \frac{c}{2
\pi } \int  \overline{L_{int}^2\tilde U_i^2}  {\cal D}^0_{l(u)}(q)dq=
-i c \frac{\overline{L_{int}^2 \tilde U_i^2}}{2 v_{II(I)}^2}
|\omega|.  \eeq
 Here $\tilde U_i$ is the  Fourier transform of the
potentials $U_i$, the bar means averaging over the scatterers   and
$c$ denotes their concentration. Thus the averaged
propagators are
 \beq \label{Eq:Dqw} {\cal  D}_{u}(q,\omega)  =  \frac
    {q}  {\omega-  v_Iq + i {v_I q}/{|\omega| \tau_u}}, \eeq
 where ${\tau_{u}(\omega)}^{-1}=c({\overline{L_{int}^2\tilde U_i^2}} /
{v_{II}^2v_I}) \omega^2  $, and a similar expression for ${\cal
D}_l$. The result for ${\tau}^{-1}$ is consistent with Eq.
(\ref{Eq:oscillations}) in the  long wavelength limit ($ 2 \pi
 v/\omega \gg L_{int}$). The obtained $\omega^2$ dependence of the
scattering rate is a consequence of the general properties of
wave scattering in continuous mediums. When the space quantization
is essential and the level spacing becomes larger than $\tau^{-1}$,
 the integration in Eq. (\ref{Eq:sigma}) should be substituted by
 summation over discrete momenta. In that case the question  of the
 symmetry between  the boundaries becomes delicate. For symmetrical
 boundaries a self-consistent treatment of the resonance width gives
 $\tau^{-1} \propto |\omega|$.

The obtained result allows  us to discuss the absorption
of external electromagnetic field when the electric field is parallel
to the edges. Using the continuity equation we can relate the
absorption coefficient $\sigma$ to the retarded propagators of the
edge waves:  \beq \sigma=\lim_{q \rightarrow 0}\frac{e^2 \omega}{q^2}
Im D^R(q,\omega).  \eeq From equation (\ref{Eq:Dqw}) we find that
in the continuous case the absorption coefficient \beq
\sigma=\frac{e^2}{\omega^2} \left( \frac{v_I}{\tau_u(\omega)} + \frac
{v_{II}} {\tau_l(\omega)}\right).  \eeq  does not depend on the
frequency because of the $\omega^2$ behavior of $\tau^{-1}$.

An interesting mechanism for electron relaxation arises as a
consequence of the random spatial variation of the inter-edge
interaction. For a homogeneous system the energy and momentum
conservation forbids an emission of waves propagating opposite to a
motion of the electron. Due to spatial inhomogeneities the momentum
conservation does not restrict the decay process anymore. After
averaging over the inhomogeneities,  the expression for
$1/\tau^e(\varepsilon)$ determining the rate of emission of the edge
waves can be written as

\beq \label{Eq:decay}
1/\tau_{u(l)}^e(\varepsilon) = \int
\Sigma_{l(u)}(\omega) {\cal G}_{u(l)}(\varepsilon-\omega,q)dqd\omega.
  \eeq
Here the integration with respect to the momentum variable $q$ in the
electron Green's function $\cal G$ should be performed independently
of $\Sigma$. This yields $1/\tau^e(\varepsilon) \propto
\varepsilon^2$ (or $\propto T^2$ at finite temperatures).
The time $\tau^e$ determines the rate of the equilibration of
electron states on the opposite edges.

In summary, we have discussed the effects related with the inter-edge
electron interaction in the quantum Hall regime. We have shown, that
experiments on propagation of the collective excitations in samples with
constrictions may give information about the amplitudes of the
electron-electron interaction which are determined by the conditions of
the screening.  We expect that the inter-edge coupling effects will be
considerably stronger in systems with a sharp transition between filled
and empty states, which were under discussions here, compared with the
systems with slowly varying electron density where electron screening
dominates \cite{QHE:Chklovskii92}.  Such effects may help to distinguish
between these two types of the quantum Hall systems.

The discussion above was related to the case when the filling factor
$\nu=1$. In case of fractional filling with $\nu=1/n$ ( $n$ is odd) the
physics of the two collective edge modes will be described, we believe, by
a phenomenological Hamiltonian $H\{\rho\}$ of the type of Eq.
(\ref{Eq:bilinear}). The specifics of the fractional state reveals in the
commutation relations for $\rho$-operators : in the right hand side of Eq
(\ref{Eq:cr}) the factor $\nu$ appears \cite{FQHE:Wen92}. This
modification does not alter the physics of the discussed phenomena.
However, certain corrections should be performed, e.g., in Eq.
(\ref{Eq:oscillations}) in the argument of the sinus a factor $\nu^{-1}$
should be introduced.

This consideration is related to wavelengths which are shorter than the
sample length. When the velocity of the edge mode is about $10^8cm/sec$ as
in the experiments of Ref.  \cite{QHE:Ashoori92,QHE:Talyanskii92}, for a
sample with length of the order of $1mm$ the frequency should be about $1
GHz$. Presumably a geometry convenient for studying the effects of the
inter-edge interaction is a pair of coupled rectangular mesas.

We thank
D.~E.~Khmelnitskii,
A.~Kamenev,
D.~Orgad   ,
N.~Argaman,
Y.~Gefen,
Y.~Imry,
Y.~Levinson,
S.~Levit and
A.~Schmid
for useful discussions.
Special acknowledgements to
U.~Meirav and
M.~I.~Reznikov
for stimulating comments.
A.F. is grateful to the Barecha Fund support.

   \bibliography{/home4/fnoreg/ref/library}

 \end{document}